\documentclass[preprint,aps,tightenlines,showpacs,nofootinbib]{revtex4}
\usepackage{latexsym}
\usepackage[dvips]{graphicx}
\newcommand{\beq}{\begin{eqnarray}}
\newcommand{\eeq}{\end{eqnarray}}
\newcommand{\eq}{eqnarray}

\newcommand{\ci}{\cite}

\newcommand{\la}{{\lambda}}
\newcommand{\La}{{\Lambda}}

\newcommand{\om}{{\omega}}
\newcommand{\Om}{{\Omega}}

\newcommand{\no}{{\nonumber}}
\newcommand{\f}{\frac}
\newcommand{\ra}{\rightarrow}

\begin{document}

\preprint{arXiv:0906.4275v2d [hep-th]}

\title{A Test of Ho\v{r}ava Gravity: The Dark Energy}

\author{Mu-In Park\footnote{E-mail address: muinpark@gmail.com}}

\affiliation{ Research Institute of Physics and Chemistry, Chonbuk
National University, Chonju 561-756, Korea }

\begin{abstract}
Recently Ho\v{r}ava proposed a renormalizable gravity theory with
higher spatial derivatives in four dimensions which reduces to
Einstein gravity with a {\it non-vanishing} cosmological constant in
IR but with improved UV behaviors. Here, I consider a non-trivial
test of the new gravity theory in FRW universe by considering an IR
modification which breaks ``softly" the detailed balance condition
in the original Ho\v{r}ava model. I separate the dark energy parts
from the usual Einstein gravity parts in the Friedman equations and
obtain the formula of the equations of state parameter.
The IR modified Ho\v{r}ava gravity seems to be consistent with the
current observational data but we need some more refined data sets
to see whether the theory is really consistent with our universe.
From the consistency of our theory, I obtain some constraints on the
allowed values of $w_0$ and $w_a$ in the Chevallier, Polarski, and
Linder's parametrization and this may be tested in the near future,
by sharpening the data sets.

\end{abstract}

\pacs{04.50.Kd, 95.36.+x, 98.80.-k }

\maketitle

\newpage

Recently Ho\v{r}ava proposed a renormalizable gravity theory with
higher spatial derivatives (up to sixth order) in four dimensions
which reduces to Einstein gravity with a {\it non-vanishing}
cosmological constant in IR but with improved UV behaviors by
abandoning Einstein's equal-footing treatment of space and time
\ci{Hora:08,Hora}. Since then various aspects
have been studied. In particular, in \ci{Lu}, it has been pointed
out that the black hole solution in the Ho\v{r}ava model does not
recover the usual Schwarzschild-AdS black hole even though the
general relativity is recovered in IR at the action level. (For
another problem in cosmology, see \ci{Nast}.) For this reason, in
\ci{Keha} an IR modification which allows the flat Minkowski vacuum
has been studied by introducing a term proportional to the Ricci
scalar of the three-geometry $\mu ^4 R^{(3)}$
(for related discussions, see also \ci{Soti}) and recently the {\it
general} black hole and cosmological solutions have been found
\ci{Park:0905}, which reduce to those of \ci{Lu} in the absence of
the IR modification term and those of \ci{Keha} for vanishing
cosmological constant ($\sim \La_W$). (For other aspects, see
\ci{Calc}).

On the other hand, in \ci{Park:0905}, the author argued that the
dark energy may be explained by the Ho\v{r}ava gravity and obtained
the equation of state parameter which seems to be consistent with
the observational data, by neglecting the matter
contributions\footnote{Recently, it has been also proposed by
Mukohyama \ci{Muko:0905_2} that the dark ``matter" as integration
constant in Ho\v{r}ava gravity, from the ``projectability" condition
\ci{Hora}. But in this paper, I consider the non-projectable case,
for the definiteness.}.

In this paper, I consider an improved analysis of the proposal by
comparing with the latest data which does not need to know about
matter contributions, separately. This would provide a possible test
of Ho\v{r}ava gravity.

To this ends, I start by considering the ADM decomposition of the
metric
\begin{\eq}
ds^2=-N^2 c^2 dt^2+g_{ij}\left(dx^i+N^i dt\right)\left(dx^j+N^j
dt\right)\
\end{\eq}
and the IR-modified Ho\v{r}ava gravity action which reads
\begin{\eq}
S_{g} &= & \int dt d^3 x
\sqrt{g}N\left[\frac{2}{\kappa^2}\left(K_{ij}K^{ij}-\lambda
K^2\right)-\frac{\kappa^2}{2\nu^4}C_{ij}C^{ij}+\frac{\kappa^2
\mu}{2\nu^2}\epsilon^{ijk} R^{(3)}_{i\ell} \nabla_{j}R^{(3)\ell}{}_k
\right.
\nonumber \\
&&\left. -\frac{\kappa^2\mu^2}{8} R^{(3)}_{ij}
R^{(3)ij}+\frac{\kappa^2 \mu^2}{8(3\lambda-1)}
\left(\frac{4\lambda-1}{4}(R^{(3)})^2-\Lambda_W R^{(3)}+3
\Lambda_W^2\right)+\frac{\kappa^2 \mu^2 \om}{8(3\lambda-1)}
R^{(3)}\right]\ , \label{horava}
\end{\eq}
where
\begin{\eq}
 K_{ij}=\frac{1}{2N}\left(\dot{g}_{ij}-\nabla_i
N_j-\nabla_jN_i\right)\
 \end{\eq}
is the extrinsic curvature (the dot $(\dot{~})$ denotes the
derivative with respect to $t$),
\begin{\eq}
 C^{ij}=\epsilon^{ik\ell}\nabla_k
\left(R^{(3)j}{}_\ell-\frac{1}{4}R^{(3)} \delta^j_\ell\right)\
 \end{\eq}
is the Cotton tensor,  $\kappa,\lambda,\nu,\mu, \La_W,\om$ are
constant parameters. The last term, which has been introduced in
\ci{Hora,Keha,Nast}, represents a ``soft" violation of the
``detailed balance" condition in \ci{Hora} and this modifies the IR
behaviors.

Now, in order to study the cosmological implications of the action
(\ref{horava}), I consider a homogeneous and isotropic cosmological
solution with the standard FRW form\footnote{In some literatures
(see \ci{Char,Blas:0906}, for example), it has been claimed that,
when considering perturbations around the background metric, the
theory does not have an IR limit close to general relativity due to
strongly coupled gravity fluctuations. 
This would be a very important issue for the consistency of the
theory though the detailed discussions about this issue (see
\ci{Gao:0905} for the troubles in \ci{Char}) is beyond the scope of
the present work.
But, in practice, this issue might not be quite
relevant to our case since there would be a natural low momentum
cut-off $\sim \sqrt{|\Lambda_W|}$
when considering fluctuations around cosmological solutions with a
non-vanishing but tiny cosmological constant $\sim \Lambda_W$, as
favored by the current observational data. This fact may change in
our non-relativistic case since the meaning of the horizons would be
also changed from the conventional ones, due to the momentum
dependence of the light cones in UV. But in IR, i.e., low momentum,
the usual meaning of the horizons would be ``emerged" from the
recovered Lorentz invariance (with $\la=1$) and so does the notion
of the low momentum cut-off. Of course, we need some more rigorous
analysis for a more explicit confirmation of this fact.
}
\begin{\eq}
ds^2=-c^2dt^2+a^2(t)\left[\frac{dr^2}{1-kr^2/R_0^2}+r^2\left(d\theta^2+\sin^2\theta
d\phi^2\right)\right],
\end{\eq}
where $k=+1,0,-1$ correspond to a closed, flat, and open universe,
respectively, and $R_0$ is the radius of spatial curvature of the
universe in the current epoch. Assuming the matter contribution to
be of the form of a perfect fluid with the energy density $\rho$ and
pressure
$p$, I find that \ci{Park:0905}
\begin{\eq}
\left(\f{\dot{a}}{a}\right)^2&=&\frac{\kappa^2}{6(3\lambda-1)}
\left[\rho \pm \frac{3\kappa^2\mu^2}{8(3\lambda-1)} \left( \f{-
k^2}{R_0^4 a^4}+ \f{2 k (\La_W -\om)}{R_0^2 a^2}- \La_W^2 \right) \right] ,  \label{F1}\\
\f{\ddot{a}}{a}&=&\frac{\kappa^2}{6(3\lambda-1)} \left[-\f{1}{2}
(\rho+3 p) \pm \frac{3 \kappa^2\mu^2}{8(3\lambda-1)} \left( \f{
k^2}{R_0^4 a^4}- \La_W^2 \right) \right], \label{F2}
\end{\eq}
where the prime $(')$ denotes the derivative with respect to $r$. I
have considered the analytic continuation $\mu^2 \ra - \mu^2$ for
$\La_W>0$ \ci{Lu,Park:0905,Park:07} and the upper
(lower) sign denotes the
$\La_W<0~ (\La_W>0)$ case. Note that the $1/a^4$ term, which is the
contribution from the higher-derivative terms in the action
(\ref{horava}), exists only for $k \neq 0$ and become dominant for
small $a$, implying that the cosmological solutions of general
relativity are recovered at large scales. The first Friedman
equation (\ref{F1}) generalizes those of \ci{Lu} and \ci{Keha} to
the case with an arbitrary cosmological constant and the soft IR
modification term in \ci{Hora,Keha,Nast}. However, it is interesting
to note that there is {\it no} contribution from the soft IR
modification to the second Friedman equation (\ref{F2}) and this is
identical to that of \ci{Lu}.

If the Friedman equations (\ref{F1}) and (\ref{F2}) are compared
with those our universe, expressed in the usual languages of the
Einstein gravity with the ``unknown'' contributions of ``dark
energy" \footnote{I follow the physical convention of Ryden
\ci{Ryde} which disagrees with \ci{Hora:08,Hora}: $G_{\rm
Here}=G_{\rm Horava}/c^3,~\La_{\rm Here}=\La_{\rm Horava} c^2$.},
\begin{\eq}
\left(\f{\dot{a}}{a}\right)^2&=&\frac{8 \pi G }{3 c^2}
(\rho_{\rm matter} +\rho_{\rm D.E.} )-\f{c^2  {k} }{R_0^2 a^2},  \label{FF1}\\
\f{\ddot{a}}{a}&=&-\frac{4 \pi G }{3 c^2} [(\rho_{\rm matter} +
\rho_{\rm D.E.})+ 3 (p_{\rm matter} + p_{\rm D. E.})],\label{FF2}
\end{\eq}
the energy density and pressure of the dark energy part can be read
as (for a related discussion with matters in the context of the
original Ho\v{r}ava gravity, see \ci{Sari})
\begin{\eq}
\rho_{\rm D.E.}&=&\pm \frac{3\kappa^2\mu^2}{8(3\lambda-1)} \left(
\f{-
k^2}{R_0^4 a^4}- \f{2 k \om}{R_0^2 a^2}- \La_W^2 \right), \label{rho}\\
p_{\rm D.E.}&=&\mp \frac{\kappa^2\mu^2}{8(3\lambda-1)} \left( \f{
k^2}{R_0^4 a^4}- \f{2 k \om}{R_0^2 a^2}-3 \La_W^2 \right),
\label{pressure}
\end{\eq}
respectively, where I have ``defined" the fundamental constants of
the speed of light $c$, the Newton's constant $G$, and the
cosmological constant $\La$ as
\begin{\eq}
c^2 = \f{ \kappa^4 \mu^2 |\La_W|}{ 8 (3 \la-1)^2},~G=\f{\kappa^2
c^2}{16 \pi (3 \la -1)}, ~{\La}=\f{3}{2} \La_{W} c^2.
\label{constant}
\end{\eq}
Note that $c^2$ is non-negative always\footnote{One might include
$\om$ term in the definition of $c^2$, rather than including in
$\rho_{\rm D.E.}$, by replacing $|\La_W| \ra |\La_W|\pm \om$ but
then $c^2$ can be negative when $|\om| > |\La_W|$.  This is what has
been considered in \ci{Keha} to get the non-vanishing speed of light
for the flat universe limit $\La_W \ra 0$ with $\omega=\pm 8
\mu^2(3\lambda-1)/\kappa^2$, though our current universe seems to
have a non-vanishing (positive) $\La_W$. But actually, there are
infinitely many possible definitions of $c^2$, depending on how much
the $\om$ term contributes to $c^2$. In this paper, however, I do
not consider all these possibilities and but consider only the
simplest choice which is related to that of the Ho\v{r}ava's
original proposal \ci{Hora,Sari}. But this would be justified by
experiments basically.}, whereas $G$ can be negative, i.e., {\it
anti}-gravity, for $\la <1/3$, which implies that $\la_{\rm c}=1/3$
is the lower bound for the consistency with our universe. Moreover,
with these definitions, one obtain the IR limit of the action
(\ref{horava}) as the sum of the $\la$-deformed Einstein-Hilbert
action $S_{\la \rm EH}$ and the IR limit of the dark energy action
with
\begin{\eq}
S_{\la \rm EH} &= & \f{c^4}{8 \pi G (3 \la-1)} \int dt d^3 x
\sqrt{g}N\left[\frac{1}{c^2}\left(K_{ij}K^{ij}-\lambda
K^2\right)+R^{(3)}-\f{2 \Lambda}{c^2} \right]\ , \label{lEH} \\
S_{\rm D.E.(IR)} &= & \f{3 c^6 \om }{32 \pi G |\La|} \int dt d^3 x
\sqrt{g}N R^{(3)}. \label{de:IR}
\end{\eq}
The deformed action $S_{\la \rm EH}$ agrees with the
Einstein-Hilbert action when $\la=1$, but otherwise there are
explicit $\la$-dependances, generally\footnote{In \ci{Hora:08,Hora}
and almost all other later works, $S_{\la \rm EH}$ is given by $
S_{\la \rm EH} =  \f{c^4}{16 \pi G } \int dt d^3 x \sqrt{g}N
[\frac{1}{c^2}(K_{ij}K^{ij}-\lambda K^2)+R^{(3)}-\f{2 \Lambda}{c^2}
] $ with $ c^2 \equiv \f{ \kappa^2 \mu^2 |\La_W|}{ 4 (3
\la-1)},~G=\f{\kappa^2 c^2}{32 \pi}, ~{\La}=\f{3}{2} \La_{W} c^2$,
in the convention of this paper \ci{Ryde}. This agrees with
(\ref{lEH}) for $\la=1$ but $\la <1/3$ is excluded from a
mathematical consistency of $c^2>0$, rather than the physical reason
of {\it no} anti-gravity, i.e., $G>0$ in the definitions of
(\ref{constant}).}. This is in contrast to the directly measurable
equations (\ref{FF1}), (\ref{FF2}) which define $c$ and $G$,
independently of $\la$; actually, the Newton's constant in
(\ref{FF1}) agrees with what we measure in laboratory, based on the
Newton's Law of Gravity, $F=-G M m/R^2$ \ci{Ryde} and we can observe
only the ``renormalized'' Newton's constant $G$, in contrast to
\ci{Keha}.

Once the energy density and pressure of dark energy (\ref{rho}),
(\ref{pressure}) are identified, one can now compute the equation of
state parameter as
\begin{\eq}
w_{\rm D.E.}=\f{p_{\rm D.E.}}{\rho_{\rm D.E.}}=\left( \f{k^2- 2 k
\bar{\om}  a^2-3 \bar{\La}_W^2 a^4}{3k^2+ 6 k \bar{\om}  a^2+3
\bar{\La}_W^2  a^4} \right) \label{EOS},
\end{\eq}
where I have introduced $\bar{\om} \equiv \om
R_0^2,~\bar{\La}_W=\La_W R_0^2$ for the convenience. This
interpolates from $w_{\rm D.E.}=1/3$ in the UV limit to $w_{\rm
D.E.}=-1$ in the IR limit but the detailed evolution patten in
between them depends on the parameters $k, \bar{\om}, \bar{\La}_W$.
There are infinite discontinuities when $\bar{\om}^2 \geq
\bar{\La}_W^2,~ k \bar{\om} <0$ due to the vanishing $\rho_{\rm
D.E.}$ with a non-vanishing $p_{\rm D.E.}$ \ci{Park:0905}. But
physically more interesting case would be $\bar{\om}^2 <
\bar{\La}_W^2,~ k \bar{\om} <0$ or $k \bar{\om} >0$ where there is
no singular point of vanishing $\rho_{\rm D.E.}$ but smoothly
fluctuating/varying between the UV and IR limits (Fig.1,2). For the
original Ho\v{r}ava gravity with $\bar{\om}=0$, $w_{\rm D.E.}$ is
``always" monotonically decreasing from $1/3$ in the UV limit to
$-1$ in the IR limit \ci{Sari}.
\begin{figure}
\includegraphics[width=10cm,keepaspectratio]{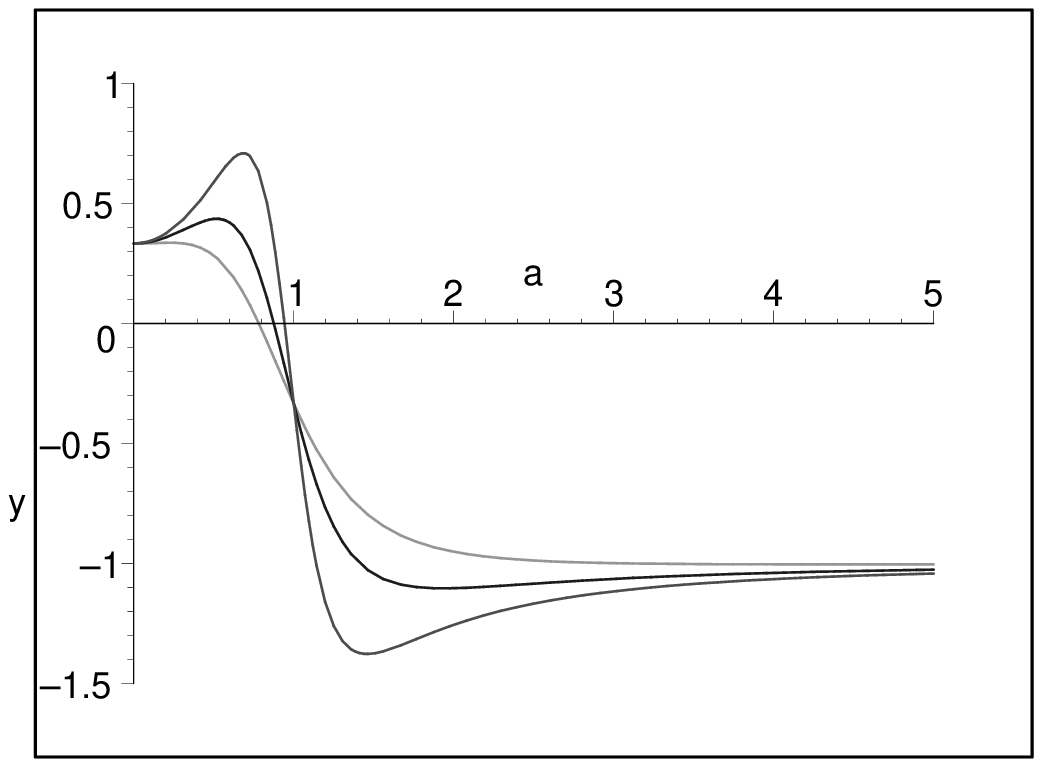}
\caption{Plots of equation of state parameters $w_{\rm D.E.}$ vs.
scale factor $a(t)$ for $\bar{\om}^2 < \bar{\La}_W^2$, $k
\bar{\om}<0$ ( $\bar{\om} =-1/1.3, -1/2, -1/10$, $k=+1$, or
$\bar{\om} =+1/1.3, +1/2, +1/10$, $k=-1$, with $|\bar{\La}_W| =1$
(top to bottom in the left region )). When $|\bar{\om}|$ is not far
from $|\bar{\La}_W|$, there is a region where $w_{\rm D.E.}$ is
fluctuating beyond the UV and IR limits. When $|\bar{\om}|$ is small
enough, $w_{\rm D.E.}$ is monotonically decreasing from $1/3$ in the
UV limit to $-1$ in the IR limit.  } \label{fig:EOS_open_Regular}
\end{figure}

\begin{figure}
\includegraphics[width=10cm,keepaspectratio]{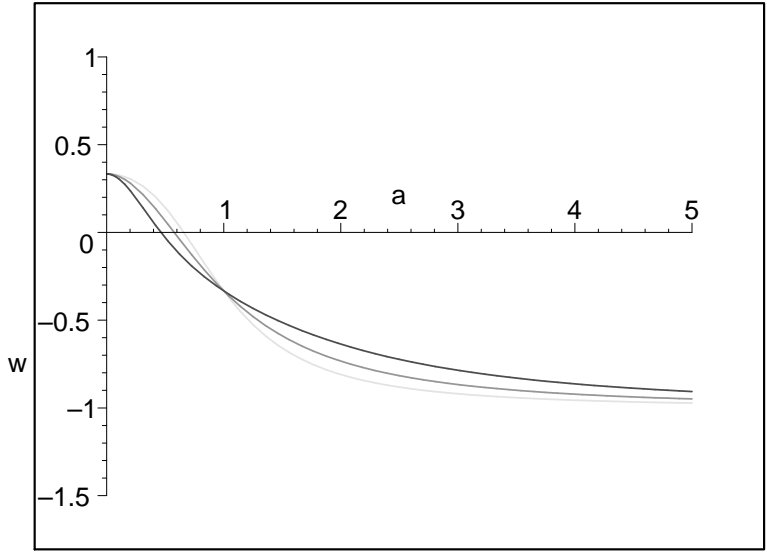}
\caption{Plots of equation of state parameters $w_{\rm D.E.}$ vs.
scale factor $a(t)$ for
$k \bar{\om}>0$ ( $\bar{\om} =\pm2, \pm 1, \pm 1/2$ (top to bottom
in the left region) for $k=\pm 1$, $|\bar{\La}_W| =1$. In this case,
$w_{\rm D.E.}$ is ``always" monotonically decreasing from $1/3$ in
the UV limit to $-1$ in the IR limit.   }
\label{fig:EOS_closed_Regular}
\end{figure}

In order to determine $w_{\rm D.E.}$, I need to know about the
constant parameters $k, \bar{\om}, \bar{\La}_W$. Previously, those
have been obtained by neglecting the matter contributions
\ci{Park:0905}. Here, I consider the more improved analysis which
does not need the consideration of the matters separately, based on
the latest observational data. To this end, let me consider the
series expansion of $w_{\rm D.E.}$ in (\ref{EOS}) near the current
epoch $(a=1)$, which coincides with Chevallier, Polarski, and
Linder's parametrization exactly \ci{Lind:02}, as
\begin{\eq}
w_{\rm D.E.}=w_0 +w_a (1-a) +w_b (1-a)^2 + \cdots \label{Linder}
\end{\eq}
with
\begin{\eq}
w_0=\f{k^2 -2 k \bar{\om}-3 \bar{\La}_W^2}{3(k^2 +2 k \bar{\om}
+\bar{\La}_W^2)}, ~ w_a=\f{8k( \bar{\om} k^2 +\bar{\om}
\bar{\La}_W^2+2 k \bar{\La}_W^2)}{3(k^2 +2 k \bar{\om}
+\bar{\La}_W^2)^2}.
\end{\eq}
So, by knowing $w_0$ and $w_a$ from the observational data, one can
determine the constant parameters $\bar{\om}$ and $\bar{\La}_W$ as
\begin{\eq}
\bar{\om}=\f{(1-2 w_0-3 w_0^2 -w_a)k}{(1+4 w_0+3 w_0^2 +w_a)},\no
\\ \bar{\La}_W^2=\f{(-1+9 w_0^2 +3w_a)k^2}{3(1+4 w_0+3 w_0^2 +w_a)}.
\label{Lambda}
\end{\eq}
Note that here I do not need to know about the matter contributions
separately, in contrast to the previous analysis \ci{Park:0905} in
which I have neglected the matter contributions to get the
approximate value of $\bar{\La}_W$ from the transition point from
deceleration phase to acceleration phase $a_T$ in (\ref{F2}). From
the latest data sets (i.e., central values of the best fits) when a
non-flat universe is allowed in the analyses \ci{Ichi,Xia}
$(w_0,w_a)=(-1.10,0.39),~(-1.06,0.72),~(-1.11,0.475)$, I get
$(\bar{\om}, \bar{\La}_W)=(1.32,2.44),~(1.14,2.10),~(1.30,2.29)$,
respectively and $k=-1$. Once the two constant parameters
$\bar{\om},\bar{\La}_W$ are determined, the whole function $w_{\rm
D.E.}(a)$ is ``completely" determined. From the obtained data, I
plot the curves of $w_{\rm D.E.}(z)$ v.s. the astronomer's variable
of redshift $z=1/a-1$ in Fig.3 and these correspond to those of
Fig.1 since $|\bar{\om}| <\bar{\La}_W$.
\begin{figure}
\includegraphics[width=10cm,keepaspectratio]{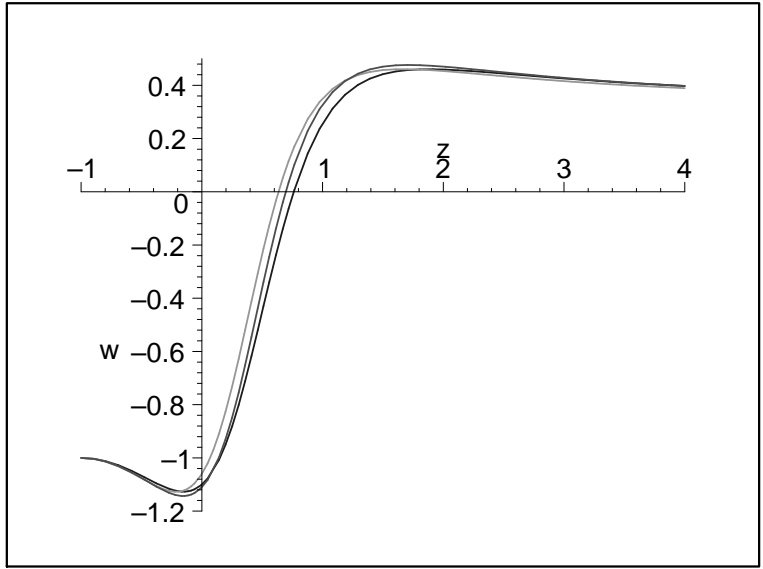}
\caption{Plots of equation of state parameters $w_{\rm D.E.}$ vs.
redshift $z=1/a-1$ for the latest data sets
$(\bar{\om},\bar{\La}_W)=(1.32, 2.44),~(1.14, 2.10),~(1.30, 2.29)$
from $(\om_0, \om_a)=(-1.10, 0.39),~(-1.06, 0.72),~ (-1.11, 0.475)$
(bottom to top) and $k=-1$. } \label{fig:EOS_closed_Regular}
\end{figure}
It is interesting to note that these curves appear to give similar
results with a nearly model independent analysis of type Ia
supernovae (SNe Ia) for $0 \leq z \leq 0.6$ \ci{Hute} and similar
tendencies in other analyses of Gold data sets that use the
parametrization (\ref{Linder}) even for higher redshifts \ci{Alam}.
\footnote{In these analyses, the flat universe has been assumed but
the results would be almost the same even for a non-flat universe.
This is due to the fact that the relaxation of flatness broadens the
ranges of $w_0$ and $w_a$ but the central values are almost
unchanged \ci{Ichi,Xia,Wang:07,Wang:08}.} In addition,
(\ref{Lambda}) gives some constraints on the allowed values of $w_0$
and $w_a$ such that $\bar{\La}_W^2 \geq 0$ for the consistency of
our theory, $w_a > \f{1}{3} (1-9 w_0^2),~-1-4 w_0-3 w_0^2$ or $w_a <
\f{1}{3} (1-9 w_0^2),~-1-4 w_0-3 w_0^2$ (See Fig. 4). This seems to
agree with observational data at about $1 \sigma~ (68.3 \%)$
confidence level \ci{Ichi,Wang:07,Alam,Upad}. This consistency
condition may be tested in the near future, by sharpening the data
sets.
\begin{figure}
\includegraphics[width=10cm,keepaspectratio]{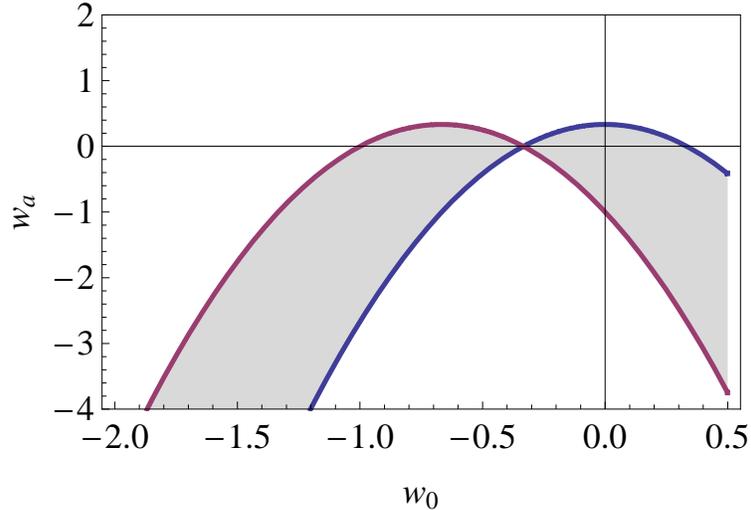}
\caption{The ranges of allowed $w_0$ and $w_a$ (unshaded regions)
for the consistency of our theory with (\ref{Lambda}). }
\end{figure}

On the other hand, using the corresponding data $\Omega_k = -0.009,
-0.000,-0.0008$ \ci{Ichi,Xia} in the current epoch $(a=1)$ for the
deviation from the critical density, $\Omega_k\equiv
1-\Om_m-\Om_{\rm D. E.}=\mu^2 k |\Lambda_W| L_{\rm P}^2/2a^2 H^2
R_0^2 M_{\rm P}^2$, Hubble parameter $H\equiv \dot{a} /a$, the ratio
of Planck mass and length $M_{\rm P}/L_{\rm P} =c^2/8 \pi G$, I get
$\mu =0.0013, 0.000, 0.0004~ (H_0 R_0 M_{\rm P}/L_{\rm P})$ with the
current value of Hubble parameter $H_0$. (See Table 1 for a summary
of the data sets and their corresponding constant parameters, in the
appropriate units.)
\begin{table}
\begin{center}
\begin{tabular}{|c|c|c|c|} \hline
Parameters at $a=1$ & Data analysis Ia \ci{Ichi}&
Data analysis Ib \ci{Ichi}& Data analysis II \ci{Xia} \\
\hline
\hline $w_0$  & -1.10 & -1.06 & -1.11 \\
\hline $w_a$  & 0.39 & 0.72 & 0.475 \\
\hline $\Om_k$  & -0.009  & -0.000 & -0.0008 \\
\hline $\Om_{\rm D.E.}$ & 0.730 & 0.699 & 0.739   \\
\hline $\Om_{\rm m}$  & 0.279 & 0.301 & 0.262 \\
\hline $H_0$  & 67.6 & 65.5 & 72.4 \\
\hline $\bar{\om}$  & 1.32 & 1.14 & 1.30 \\
\hline $\bar{\La}_W$ & 2.44 & 2.10 & 2.29 \\
\hline $\bar{\om} /\bar{\La}_W$ & 0.542  & 0.543 & 0.568   \\
\hline $\mu$  & 0.0013 & 0.0000 &0.0004\\
\hline
\end{tabular}
\end{center}
\caption{A summary of the data sets {\it without} assuming the flat
universe in a priori and their corresponding constant parameters, in
the conventional units of $H_0~ ({\rm km ~s^{-1} Mpc^{-1}})$ and
$\mu ~(H_0 R_0 M_{\rm P}/L_{\rm P})$.}
\end{table}

In conclusion, I have considered the dark energy as a possible test
of Ho\v{r}ava gravity. It seems that the IR modified Ho\v{r}ava
gravity seems to be consistent with the current observational data
but we need some more refined data sets to see
whether the theory is really consistent with our universe. However,
it would be still useful to consider the list of possible scenarios
as follows.

1. If $k=0$, i.e., flat universe, as predicted by inflationary
cosmology but {\it not} compulsory in the latest analyses
\ci{Ichi,Xia,Wang:07,Wang:08}, is confirmed, there is no effect of
the Ho\v{r}ava gravity in the FRW cosmology. But even in this case,
its effect to the anisotropic cosmology and non-Gaussianity would be
still open problems.

2. If $w_{\rm D.E.}<-1$ and $k \neq 0$, the original Ho\v{r}ava
gravity with the detailed balance condition, which predicts $-1 \leq
w_{\rm D.E.} \leq 1/3$, may be ruled out.  According to the current
observational data, this scenario seems to be quite plausible and
this is also consistent with other theoretical considerations
\ci{Lu,Nast,Keha,Park:0905}.

3. Even if $w_{\rm D.E.}<-1$, $k \neq 0$, and good agreements for
small $z$ are confirmed by determining the constant parameters $\om$
and $\La_W$, some disagreements or inconsistencies for higher $z$
can occur; for example, by comparing the higher-order term $w_b$ in
(\ref{Linder}) with the experiments. In this case, one might
consider several further modifications of the Ho\v{r}ava gravity by
introducing other detailed-balance breaking terms with the
additional constant parameters to control the disagreements. But one
does not know how much new terms are needed minimally, in a priori
\footnote{For a related discussion, see also \ci{Lind:05}.}. Or, one
might consider another definition of $\rho_{\rm D.E.}$ and $p_{\rm
D.E.}$, by considering different definitions of the speed of light,
rather than the simplest choice (\ref{constant}), as was discussed
in the footnote no.1.

\section*{Acknowledgments}

I would like to thank  Yungui Gong, Qing-Guo Huang, Jean Iliopoulos,
Yoonbai Kim, Bum-Hoon Lee, Eric Linder, Shinji Mukohyama, Horatiu
Nastase, Sang-Jin Sin, Hyunseok Yang for helpful comments,
discussions, and also John J. Oh, Chul-Moon Yoo for technical helps.
This work was supported by the Korea Research Foundation Grant
funded by Korea Government(MOEHRD) (KRF-2007-359-C00011).

\newcommand{\J}[4]{#1 {\bf #2} #3 (#4)}
\newcommand{\andJ}[3]{{\bf #1} (#2) #3}
\newcommand{\AP}{Ann. Phys. (N.Y.)}
\newcommand{\MPL}{Mod. Phys. Lett.}
\newcommand{\NP}{Nucl. Phys.}
\newcommand{\PL}{Phys. Lett.}
\newcommand{\PR}{Phys. Rev. D}
\newcommand{\PRL}{Phys. Rev. Lett.}
\newcommand{\PTP}{Prog. Theor. Phys.}
\newcommand{\hep}[1]{ hep-th/{#1}}
\newcommand{\hepp}[1]{ hep-ph/{#1}}
\newcommand{\hepg}[1]{ gr-qc/{#1}}
\newcommand{\bi}{ \bibitem}

\end{document}